\documentclass{article}

\catcode`\@=11
\@addtoreset{equation}{section}

\global\arraycolsep=1pt
\def\fnote#1#2{\begingroup\def\thefootnote{#1}\footnote{#2}\addtocounter
{footnote}{-1}\endgroup}

\usepackage{graphicx}
\usepackage{amsmath,amssymb,amsbsy,latexsym}

\def\CC{{\mathcal C}}\def\CG{{\mathcal G}}\def\CN{{\mathcal N}}\def\CW{{\mathcal W}}\def\CZ{{\mathcal Z}}

\def\BR{{\mathbb R}}\def\BZ{{\mathbb Z}}

\def\FA{{\mathfrak A}}\def\FB{{\mathfrak B}}\def\FC{{\mathfrak C}}\def\FD{{\mathfrak D}}\def\FS{{\mathfrak S}}

\def\tr{{\rm Tr}}

\def\vol{{\rm vol}}
\def\res{{\rm Res}}

\def\divar#1#2#3{\{ #1 \stackrel{#3}{\Longleftrightarrow} #2\}}
\def\divbr#1#2#3{\{ #1 \stackrel{#3}{\mbox{- - - -}} #2\}}
\def\infrac#1#2{\displaystyle\frac{#1}{#2}}

\def\SU#1{{\rm SU}{$(#1)$}}
\def\SO#1{{\rm SO}{$(#1)$}}
\def\USp#1{{\rm USp}{$(#1)$}}

\def\mSU#1{{\rm SU}{(#1)}}
\def\mSO#1{{\rm SO}{(#1)}}
\def\mUSp#1{{\rm USp}{(#1)}}

\begin{document}

\begin{flushright}
OCU-PHYS-253\\
September 2006\\
hep-th/0609063
\end{flushright}
\bigskip

\begin{center}
{\bf\Large
Partition Functions of Reduced Matrix Models \\
with Classical Gauge Groups
}

\bigskip\bigskip
{
H. Itoyama$^{a,b}$\fnote{$*$}{e-mail: \texttt{itoyama@sci.osaka-cu.ac.jp}}
\quad, \quad
H. Kihara$^b$\fnote{$\dag$}{e-mail: \texttt{kihara@sci.osaka-cu.ac.jp}}
\quad and \quad
R. Yoshioka$^a$\fnote{$\ddagger$}{e-mail: \texttt{yoshioka@sci.osaka-cu.ac.jp}}
}
\end{center}

\bigskip

\begin{center}
$^a$ {\it Department of Mathematics and Physics,
Graduate School of Science\\
Osaka City University\\
3-3-138, Sugimoto, Sumiyoshi-ku, Osaka, 558-8585, Japan }

\bigskip

$^b$ {\it Osaka City University, Advanced Mathematical Institute (OCAMI),\\
3-3-138 Sugimoto, Sumiyoshi, Osaka 558-8585, Japan}

\end{center}

\vfill

\begin{abstract}
We evaluate partition functions of matrix models which are given by
 topologically twisted and dimensionally reduced actions of $d=4$ $\CN=1$ super Yang-Mills theories with classical (semi-)simple gauge groups, \SO{2N}, \SO{2N+1} and \USp{2N}. 
The integrals reduce to those over the maximal tori by semi-classical approximation which is exact in reduced models. 
We carry out residue calculus by developing a diagrammatic method, 
 in which the action of the Weyl groups and therefore counting of multiplicities are explained obviously. 
 \end{abstract}

\section{Introduction}

\hspace{12pt} In 1996 Banks, Fischler, Shenker and Susskind (abbreviated BFSS) suggested 
the equivalence between 11-dimensional M-theory and the $N \rightarrow \infty$ limit 
of the supersymmetric matrix quantum mechanics describing D0-branes \cite{Banks:1996vh}. 
The action of their model is obtained by the reduction of 
$d=10 ~~\CN=1$ super Yang-Mills theory with gauge group \SU{N} \cite{deWit:1988ig}.
Ishibashi, Kawai, Kitazawa and Tsuchiya (IKKT) proposed a zero-dimensional matrix model 
with manifest ten-dimensional $\CN=2$ super Poincar\'e invariance \cite{Ishibashi:1996xs}.
The action of their model is given by reduction to zero dimension
 of the $\CN =1, d=10$ super
 Yang-Mills action with gauge group $G=SU(N)/{\BZ }_N$.
We will call it IKKT action.
Hirano and Kato showed that the IKKT action is topological \cite{Hirano:1997ai}.
 Topological field theory is introduced in \cite{Witten:1988ze}.
In 1998 Moore, Nekrasov and Shatashvili (MNS) \cite{Moore:1998et}, 
being motivated by the existence of D-particle bound states \cite{Witten:1995im,Yi:1997eg,Sethi:1997pa,Porrati:1997ej,Green:1997tn},
 computed the partition functions of the zero-dimensional supersymmetric matrix models 
 as the deficit terms of the Witten indices \cite{Witten:1981nf,witind,vanBaal:2001ac},  
 \fnote{\dag 1}{van Baal attempted to deal with the orbifold singularities
 in the moduli space of flat connections for supersymmetric gauge theories
 on the torus. The vacuum valley parametrized by the abelian zero-momentum modes
 and the effective Hamiltonian requires modification due to a singularity 
in the non-adiabatic behavior at the orbifold singularities. }

\begin{align}
\CZ &=  \frac{1}{\vol(G)} \int [dX][d\Psi] e^{-S}.
\label{eqn:1}
\end{align}
Here $S$ is the IKKT action 
and we denote bosonic matrices by $X$ and fermionic matrices by $\Psi$ collectively.

The partition functions of matrix models are expressed as functional integrals with
 the actions reduced from higher (4,6 and 10) dimensional gauge theories to their 
 zero-dimensional counterparts.
 MNS treated the topologically twisted models. 
The action given by reduction to zero dimension
 of the topologically twisted $\CN =1, d$-dimensional super
 Yang-Mills action with gauge group $G$ is denoted by MNS($d,G$) action.
MNS obtained the value of the integral for the MNS($4,$\SU{N}$/\BZ_N$) action
 by using the Cauchy determinant formula. 
They also obtained the results for $d=6$ and $d=10$ with gauge group $G=$\SU{N}$/\BZ_N$.
Kostov {\sl et. al.} studied the partition functions or the correlation functions
 of the models reduced to various dimensions \cite{Kazakov:1998ji,Kostov:1998pg}. 
Suyama and Tsuchiya calculated the exact partition function of the IIB matrix model
 with gauge group \SU{2} \cite{Suyama:1997ig}. 
Sugino {\sl et. al.} have developed the improved Gaussian and mean field approximation method for the reduced Yang-Mills integrals  
\cite{Oda:2000im}.
Austing and Wheater discussed the finiteness of the 
 SU(N) bosonic Yang-Mills matrix integrals  \cite{Austing:2001bd}. 
Dorey {\sl et. al.} claimed that in a certain limit the D-instanton partition function reduces to the functional integral of $\CN=4$ U(N) supersymmetric gauge theory for multi-instanton solutions \cite{Dorey:2000zq}. Their review on the calculus of many instantons is helpful. 

\section{Preliminaries}

Generalizations of equation (\ref{eqn:1}) to  orthogonal and symplectic groups are discussed
 and partial results are obtained \cite{Krauth:1998xh,Kac:1999gw,Pestun:2002rr}.
In this article we evaluate matrix integrals for the MNS($4,G$) actions
 in cases of $G=$  \SO{2N}, \SO{2N+1} and \USp{2N}.

We use the following notations. 
Let $G$ be a Lie group.
 $\CG$ is the Lie algebra of G and $\CC \subset \CG$ is a
Cartan subalgebra of $\CG$. Below, we list some examples of Cartan subalgebras for classical groups. 
\begin{enumerate}
\item SU(N) : $\CC=\{  i\phi ; \phi=\text{diag}(\phi_1,\cdots,\phi_N), 
\phi_1+\cdots+\phi_N=0, \phi_i\in\BR \}$
\item SO(2N) : $\CC=\{ \phi; \phi=\phi_1 J \oplus \cdots \oplus \phi_N J, 
\phi_i\in\BR \}$
\item USp(2N) : $\CC=\{ \phi; \phi=\phi_1 J \oplus \cdots \oplus \phi_N J, 
\phi_i\in\BR \}$
\item SO(2N+1) : $\CC=\{ \phi; \phi=\phi_1 J \oplus \cdots \oplus \phi_N J\oplus (0), 
\phi_i\in\BR \}$
\end{enumerate}
where $J= i \sigma_2$. 
Let $\Phi$ be a root system associated with $\CC$.
We denote the dual space of $\CC$ by $\CC^*$.
Let $\phi\in\CC$ and $\alpha\in\CC^*$. We define the inner product
 $\langle \alpha ,\phi\rangle = \alpha(\phi)$. 
The dual basis $e_i$ is defined by $\langle e_i,\phi\rangle =\phi_i$ for classical groups.
\begin{align}
&G=\mSU{N}~~,&&
 A_{N-1} = \{ \pm ( e_i - e_j)~, ~~( i < j)
 \},\cr
&G=\mSO{2N+1}~~,&&
 B_{N} = \{ \pm ( e_i - e_j)~,~ \pm ( e_i + e_j)~,~\pm
 e_k ~~(i < j )
 \},\cr
&G=\mUSp{2N}~~,&&
 C_{N} = \{ \pm ( e_i - e_j)~,~ \pm ( e_i + e_j)~,~\pm
 2 e_k ~~( i < j )
 \},\cr
&G=\mSO{2N}~~,&& D_{N} = \{ \pm ( e_i - e_j)~,~ \pm ( e_i + e_j)~, ~~( i < j )
 \}~~.
 \label{eqn:rootsystem}
\end{align}
In (\ref{eqn:rootsystem}) A$_N$, B$_N$, C$_N$ and D$_N$ are the root systems associated with the
  Lie algebras of SU(N+1), SO(2N+1), USp(2N) and SO(2N), respectively. 
The index $N$ is the rank of the root system. 
We denote the Weyl group by $W_{\Phi}$   and the center of the group G  by $Z_{G}$. 
We summarize the orders of $W_{\Phi}$ and $Z_{G}$ for A$_N$, B$_N$, C$_N$ and D$_N$ series in Table \ref{tbl:weylcenter}.

\begin{table}[h]
\caption{Orders of centers and  the Weyl groups for classical groups}
\begin{center}
\begin{tabular}{lcccc}
\hline
\hline
$\Phi$&$A_{N-1}$ & $B_N$ & $C_N$ & $D_N$ \\
\hline
G& \SU{N} & \SO{2N+1} &\USp{2N} & \SO{2N}\\
$\#W_{\Phi}$& $N!$ & $N! 2^N$ & $N! 2^N$ & $N! 2^{N-1}$\\
$\#Z_{G}$ & $N$ & $1$ & $2$ & $2$\\
\hline
\end{tabular}
\end{center}
\label{tbl:weylcenter}
\end{table}

 The number of set $X$ is denoted by $\# X$. 
Each root $\alpha \in \Phi$ defines a hyperplane $\langle \alpha,\phi\rangle=0$ in the vector space $\CC$. 
These hyperplanes divide the space $\CC$ into finitely many connected components called the Weyl chambers. 
These are open, convex subsets of $\CC$.

We now consider the partition function $\CZ_{\Phi}$ of the model with a gauge group $G$, where
$\Phi$ is a root system associated with $G$,
\begin{align}
\CZ_{\Phi} &=  \frac{1}{\vol(G)} \int [dX][d\Psi] e^{-S},\cr
S&=\tr\left(\frac{1}{4} [X_{M},X_{N}]^2 - \frac{1}{2} \bar{\Psi} \Gamma^M [X_M , \Psi] \right)~(M,N=1,2,3,4)~.
\end{align}
Here $X_M,\Psi$ are $\CG$-valued and the measures $[dX],[d\Psi]$ are $G$-invariant measures in this article.
We choose two matrices $X_3$ and $X_4$ and arrange them into a complex matrix $\phi=X_3+iX_4$.
According to the prescription of MNS \cite{Moore:1998et} the functional integrals reduce to 
the integral of $\phi$.  
In addition  $\phi$ can be integrated over $\CG$. 
In reducing the integral on $\phi$ from $\CG$ to $\CC$ we obtain the integral
\begin{align}
\CZ_{\Phi} &= \frac{1}{E^{r}} \frac{\# Z_{G}}{\# W_{\Phi}} \left(
 \prod_{\ell=1}^r  \int_{-\infty}^{\infty} \frac{d\phi_{\ell}}{2\pi
 \sqrt{-1}}  \right) \prod_{\alpha \in \Phi} \frac{\langle \alpha,\phi\rangle }{\langle\alpha,\phi\rangle-E}~~.
\label{eqn:intofphi} 
\end{align}
Here $E$ is a deformation parameter associated with the global symmetry SO(2) and $r$ is the rank of $\Phi$. 
The integrand is a rational function of $\phi_i$ and the degree of the denominator is equal to that
of the numerator. 
Naively the integrals diverge, so we must regularize and renormalize the integrals. 
We cut off the integrals by introducing a parameter $\Lambda$ temporarily. 
We add an integral along the upper half-circle with radius $\Lambda$ in every 
$\phi_i$ plane as a counter term. 
Then the renormalized partition function becomes
\begin{align}
\CZ_{\Phi}^{\rm R} &= \frac{1}{E^{r}} \frac{\# Z_{G}}{\# W_{\Phi}} \left(
 \prod_{\ell=1}^r  \oint \frac{d\phi_{\ell}}{2\pi
 \sqrt{-1}}  \right) \prod_{\alpha \in \Phi} \frac{\langle\alpha,\phi\rangle}{\langle\alpha,\phi\rangle-E}~~.
\label{eqn:intofphi2} 
\end{align}
We work on the residue calculations. 
We shift $E$ to the pure imaginary direction to avoid the poles.
The nontrivial contributions come from  points in which
  at least $r$ divisors  $\{ \langle\alpha_a,\phi\rangle-E\}_{a=1, \cdots , r}$ take zero value. 
Such a point is a solution of a system of linear equations
 $\{\langle\alpha,\phi\rangle -E =0\}_{\alpha \in \Lambda},
~(\Lambda \subset \Phi:\text{subset}, ~\# \Lambda \geq r)$.
The roots are separated into two kinds, positive and negative roots
 by a certain partial order.
The collections of  positive and negative roots will usually be denoted $\Phi_+$ and $\Phi_-$, respectively.
The integrand can be regarded as a function of $\langle\alpha,\phi\rangle(\alpha\in\Phi_+)$,
\begin{align}
\CZ_{\Phi}^{\rm R} &= \frac{1}{E^{r}} \frac{\# Z_{G}}{\# W_{\Phi}} \left(
 \prod_{\ell=1}^r  \oint \frac{d\phi_{\ell}}{2\pi
 \sqrt{-1}}  \right) \prod_{\alpha \in \Phi_+} \frac{\langle\alpha,\phi\rangle^2}{\langle\alpha,\phi\rangle^2-E^2}~~.
\label{eqn:intofphi3} 
\end{align}
We will carry out these integrals for A$_N$, B$_N$, C$_N$ and D$_N$. 

\section{$A_{N-1}$ series}

\hspace{12pt} Let us reproduce the result for A$_{N-1}$. 
The element $\phi$ of the Cartan subalgebra is a traceless hermitian matrix as mentioned above.
Let $\Lambda$ ($\# \Lambda=k$) be a subset of A$_{N-1}$;
 $\Lambda=\{ \alpha_a= \lambda^a (e_{i_a} - e_{j_a}),~i_a<j_a,~\lambda^a\in \{\pm1\},~a=1,2,\cdots,k\}$. 

\begin{figure}[h]
\centerline{\resizebox{120mm}{!}{\includegraphics{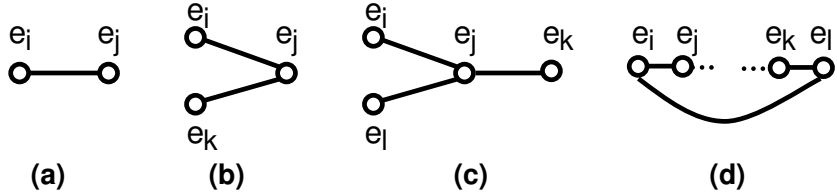}}}
\caption{(a)  root $e_i-e_j \in \Lambda$,~~(b)$\Gamma_f$: folded
 diagram,~~(c)$\Gamma_b$: branching diagram,~~(d)$\Gamma_l$: loop diagram}
 \label{fgr:diagram}
\end{figure}

We explain how to draw the diagram $\Gamma_{\Lambda}$ associated with $\Lambda$.
First of all, we draw a small circle for each $ e_i,( i=1,2,\cdots ,N)$.
  The circle for $ e_i $ is denoted by $ \{ i \} $. 
Next, we draw a line between two circles $\{i_a\}$ and $\{j_a\}$.
The line is  endowed with  the sign $\lambda^a$ for each $a=1,\dots,k$. 
The diagram $\Gamma_{\Lambda}$ consists of $N$ small circles and $k$ lines endowed with sign $\pm$.   
The line from $\{i\}$ to $\{j\}$ endowed with $\lambda$ is denoted by $\divar{i}{j}{\lambda}~(i < j)$. 
Some typical diagrams are depicted in Figure \ref{fgr:diagram}.
We obtain a system of linear equations
 from a subset $\Lambda$;
\begin{align}
\langle\alpha , \phi\rangle -E &=0~, \alpha \in \Lambda~~.
\end{align}
The line $\divar{i_a}{j_a}{\lambda^a}$ corresponds to an equation; $\lambda_a (\phi_{i_a} - \phi_{j_a}) - E=0 $. 
To evaluate the integral we consider diagrams which include just $k=N-1$ lines.
One such diagram $\Gamma_{\Lambda}$ corresponds
 to a term contributing to the partition function. 
The term is given as a residue at a zero of such a system 
 $\{\langle\alpha, \phi\rangle -E =0\}_{\alpha \in \Lambda}$. 

We can show that many of diagrams do not contribute to the partition function. 
Indeed a diagram including 
folded diagrams $\divar{i}{j}{+}$ $\divar{j}{k}{-}$
\footnote{``Folded" means that the two lines attached to the same circle are endowed with  different signs. }
, loop diagrams
$\divar{i}{j}{+} \cdots \divar{k}{l}{+}$ $\divar{i}{l}{\pm}$ and
branching diagrams $\divar{i}{j}{+}$ $\divar{j}{k}{+}$
$\divar{j}{l}{\pm}$  as subdiagrams does not contribute. 
Let us  prove this statement.
We consider three circles $\{i\},\{j\},\{k\}$ and draw a line
$\divar{i}{j}{+}$. 
Because we obtain an equation $\phi_i-\phi_j=E$ from the line, we  consider the residue at $\phi_i=\phi_j+E$,
\begin{align}
&\oint \frac{d\phi_i}{2\pi \sqrt{-1}}\oint \frac{d\phi_j}{2\pi \sqrt{-1}}\frac{(\phi_i-\phi_j)^2}{(\phi_i-\phi_j)^2 -E^2
 }\frac{(\phi_i-\phi_k)^2}{(\phi_i-\phi_k)^2 -E^2
 }\frac{(\phi_j-\phi_k)^2}{(\phi_j-\phi_k)^2 -E^2 } \cr
&\rightarrow
 \oint \frac{d\phi_j}{2\pi \sqrt{-1}}\frac{E}{2} \frac{(\phi_j-\phi_k+E)^2}{(\phi_j-\phi_k+E)^2 -E^2 }\frac{(\phi_j-\phi_k)^2}{(\phi_j-\phi_k)^2 -E^2 }\cr
&= \oint \frac{d\phi_j}{2\pi \sqrt{-1}} \frac{E}{2}
 \frac{(\phi_j-\phi_k+E)(\phi_j-\phi_k)}{(\phi_j-\phi_k+2E)(\phi_j -
 \phi_k -E)}~~.
\end{align}
The factors corresponding to $\divar{i}{k}{+}$ and 
$\divar{j}{k}{-}$ in the denominator are divided by factors in the numerator.
The factors corresponding to $\divar{j}{k}{+}$ and $\divar{i}{k}{-}$ remain. 
The factor $\phi_j-\phi_k+2E$ is obtained from the factor $\phi_i-\phi_k+E$. 
This result shows that a diagram including a folded diagram does not contribute. 

We concretely calculate the residue at the zero of $\{\phi_i-\phi_j-E=0,\phi_j-\phi_k-E=0\}$,
\begin{align}
\oint \frac{d\phi_i}{2\pi \sqrt{-1}}\oint \frac{d\phi_j}{2\pi \sqrt{-1}}\frac{(\phi_i-\phi_j)^2}{(\phi_i-\phi_j)^2 -E^2
 }\frac{(\phi_i-\phi_k)^2}{(\phi_i-\phi_k)^2 -E^2
 }\frac{(\phi_j-\phi_k)^2}{(\phi_j-\phi_k)^2 -E^2 } &\rightarrow \frac{E^2}{3}
 ~~.
 \label{eqn:loop}
\end{align} 
The factor $\phi_i-\phi_k$ is a sum of the two factors $(\phi_i-\phi_j)$ and $(\phi_j-\phi_k)$.
Thus the factor  $\phi_i - \phi_k$ is not linearly independent of $ \phi_i-\phi_j$ and $\phi_j - \phi_k$.  This argument is easily generalized to a long loop $\divar{i_1}{i_2}{+} \divar{i_2}{i_3}{+}\cdots \divar{i_{p-1}}{i_p}{+}$ $\divar{i_1}{i_p}{\pm}$. The factor $\phi_{i_1}-\phi_{i_p}$
 is a sum $\sum_{a=1}^{p-1} (\phi_{i_a}-\phi_{i_{a+1}})$. 
Thus there is no solution to $\{\phi_{i_a}-\phi_{i_{a+1}}-E=0~(a=1,\dots,p-1), \phi_{i_1}-\phi_{i_p}-E=0\}$($p>2$).
This result includes that there is no contribution from a diagram including a loop subdiagram either.

Next we consider four circles $\{i\},\{j\},\{k\}, \{l\}$ and draw two lines
$\divar{i}{j}{+}$ and $\divar{j}{k}{+}$. 
The corresponding residue calculation is as follows:
\begin{align}
&\oint \frac{d\phi_i}{2\pi \sqrt{-1}}\oint \frac{d\phi_j}{2\pi \sqrt{-1}}\oint \frac{d\phi_k}{2\pi \sqrt{-1}}
\frac{(\phi_i-\phi_j)^2}{(\phi_i-\phi_j)^2 -E^2}
\frac{(\phi_i-\phi_k)^2}{(\phi_i-\phi_k)^2 -E^2}\cr
&\times
\frac{(\phi_i-\phi_l)^2}{(\phi_i-\phi_l)^2 -E^2}
\frac{(\phi_j-\phi_k)^2}{(\phi_j-\phi_k)^2 -E^2}
\frac{(\phi_j-\phi_l)^2}{(\phi_j-\phi_l)^2 -E^2}
\frac{(\phi_k-\phi_l)^2}{(\phi_k-\phi_l)^2 -E^2}\cr
&\rightarrow \oint \frac{d\phi_k}{2\pi \sqrt{-1}} \frac{E^2}{3} \frac{(\phi_k-\phi_l+2E)(\phi_k-\phi_l)}{(\phi_k-\phi_l+3E)(\phi_k-\phi_l-E)}~.
\end{align}
Two factors corresponding to $\divar{k}{l}{+}$ and $\divar{i}{l}{-}$ remained.
We take no account of branching diagrams because of this result.  
Thus these results imply that there is no contribution from the diagrams which include 
 one of these three types as a subdiagram.

We can draw only straight line  configurations like $\divar{i}{i+1}{+}$
($i=1,2,\cdots,N-1$) by this prescription. 
In fact every allowed diagram can be  transformed into the diagram  $\divar{i}{i+1}{+}$
($i=1,2,\cdots,N-1$) with  a Weyl transformation which reorders their indices.
In addition, $\{e_i-e_{i+1},~(i=1,\cdots,N-1)\}$ is a fundamental root system.
 One might think that only  diagrams constructed from fundamental root systems
 are relevant for every gauge group.
We will show later that this inference is not correct. 
Let us continue the remaining calculation for A$_{N-1}$. 
The residue at the solution to $\{(\phi_i-\phi_{i+1})-E=0\}_{i=1,\dots,N-1}$ is explicitly calculated.
Taking account of the multiplicity $(N-1)!$ caused by the Weyl group,  
we finish the calculation of the partition function for A$_{N-1}$,
\begin{align}
\CZ_{A_{N-1}}
&= (N-1)!  \frac{1}{E^{N}} \frac{N}{N!}  \left( \frac{E}{2}
 \right)^{N-1} \frac{1}{N} \left( \prod_{i<j,j-i\neq 1}
 \frac{(z_i-z_j)^2}{(z_i-z_j)^2-E^2}   \right)\cr
&=\frac{ 1}{N^2} 
~~.\cr
z_i&= \frac{1}{2} (N+1-2i)E~.
\label{eqn:intofA} 
\end{align}
This result agrees with the original result of MNS, which is derived from the Cauchy determinant formula. 
This demonstrates that our diagrammatic method works for A$_{N-1}$.
 In the remainder of this article we carry out the same calculus for other classical groups.

\section{$D_N$ series}
\hspace{12pt}
 In this section
 we develop the diagrammatic method for the D$_N$ series properly.
The gauge group associated with the root system D$_N$ is \SO{2N}$/\BZ_2$ ($ Z_{SO(2N)}=\BZ_2$). 
The difference between \SO{2N} and \SO{2N}$/\BZ_2$ gives rise to the difference
 in the multiplicities. 
We do not take care of this difference until we consider the multiplicities. 
The root system D$_N$  
consists of roots $\{\pm(e_i-e_j)\}$ and $\{\pm(e_i+e_j) \}$. 
We evaluate the renormalized partition function $\CZ_{D_N}$ for the root system $D_{N}$.
\begin{align}
\CZ_{D_N} 
&= \frac{1}{E^{N}} \frac{2}{N!2^{N-1}} \left(
 \prod_{\ell=1}^N  \oint \frac{d\phi_{\ell}}{2\pi
 \sqrt{-1}}  \right) \prod_{\alpha \in D_N} \frac{\langle\alpha,\phi\rangle}{\langle\alpha,\phi\rangle-E}~,
\label{eqn:intofD} 
\end{align}
where we have omitted the superscript ``${\rm R}$". 
 The Weyl transformation plays an important role in this case. 
The Weyl group consists of permutations
$i\rightarrow \sigma(i), \sigma \in \FS_N$ and sign flips $(e_i,e_j)
\rightarrow (-e_i,-e_j)$ $(i<j)$. The integrand is also invariant under
sign flips $e_i \rightarrow -e_i$. 
The root system $D_{N}$ includes two kinds of positive roots $e_i-e_j$
and $e_i+e_j$ $(i<j)$.

\begin{figure}[h]
\centerline{\resizebox{50mm}{!}{\includegraphics{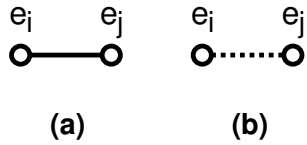}}}
\caption{(a) $e_i-e_j$,~~(b) $e_i+e_j$}
\label{fig:D}
\end{figure}

Now we extend the diagrammatic method. 
Let us draw circles for $e_i$ and lines for the roots as well as those for \SU{N}. 
To express the   
  difference between two kinds of positive roots, 
we use solid lines for $e_i-e_j$ and broken lines for
$e_i+e_j$.  
A solid line between $\{i\}$ and $\{j\}$ is represented by $\divar{i}{j}{+}$  
 and a broken line between these is by $\divbr{i}{j}{+}$ as Figure \ref{fig:D}.  
 Each diagram includes $N$ circles. 
Because the partition function has $N$ integrations, 
diagrams with $N$ lines contribute to the partition function. 
Every diagram with $N$ circles and $N$ lines must contain at least one loop subdiagram. 
The number of loops corresponds to that of connected components of the diagram.  
The symmetry induces transformations on diagrams.  
Two diagrams which can be transformed each other yield the same contribution.  
In particular every connected diagram can be transformed  into  
  a diagram with only one broken line.

Let us prove this statement partially.   
If we encounter a broken line $\divbr{i}{j}{+}$, we change the sign $e_j$.  
Then the line $\divbr{i}{j}{+}$ is transformed into $\divar{i}{j}{+}$. 
 These transformations eliminate almost all broken lines.  
However if we encounter $\divar{i}{j}{\pm}$ and $\divbr{k}{j}{\pm}$, 
 we cannot decrease the number of broken lines,  
 because the sign flip does not change the number.
Though there is no proof that disconnected diagrams do not contribute, 
 we consider only connected diagrams. 
This conjecture is partially obtained from some explicit evaluations. 
A connected diagram can include only one loop because two loops imply the disconnectedness.
For solid lines the rules in the case of $A_{N-1}$ are still valid. 
The branching $\divar{i}{j}{}$$\divar{j}{k}{}$$\divbr{j}{l}{}$ 
 cannot happen because the sign flip $e_l \to -e_l$ transforms that to the branching of solid lines.  
Thus the possible cases are $\divar{i}{j}{}$$\divar{j}{k}{}$$\divbr{j}{k}{}$. Here we omitted their signs.
Hence the connected diagram which contributes to the partition function is transformed into a diagram 
 which has only one broken line in the loop part $\divar{i}{j}{}$$\divbr{i}{j}{}$.  
 We have arrived at our result.

Now we consider a valid connected diagram with one broken line. 
Its subdiagram consisting of all of the solid
lines is a straight configuration which is the same one as $A_{N-1}$. 
It reflects that the group \SU{N} is  a subgroup of \SO{2N}.
The following change of variables makes this situation clear;
\begin{align}
&x =\rho_0= \frac{1}{N}(\phi_1+\phi_2+\cdots \phi_N)~,&
\rho_i &= \phi_i - x,\cr
&\rho_1+\rho_2 + \cdots +\rho_N =0~, & J &=
 \frac{\partial(\rho_0,\rho_1, \cdots,\rho_{N-1})}{\partial(\phi_1, \phi_2 ,\cdots, \phi_N
 )} =\frac{(-1)^{N-1}}{N}~~.
\end{align}
Here we separate coordinates into ``center of mass" $x$ and ``relative coordinates" $\rho_i$. 
Then the factors, $\langle \alpha, \phi \rangle$,  are written as, 
\begin{align}
\phi_i - \phi_j &= \rho_i - \rho_j~, & \phi_i+\phi_j &=
 \rho_i+\rho_j+2x~~.
\label{eqn:poleass}
\end{align}
We pick up residues at $\rho_i-\rho_{i+1}=+E$ ($i=1,2,\cdots,N-1$). 
This is the same system of equations as that appeared in the case of $A_{N-1}$.  
This residue calculation reduces the original integral to the one over a Weyl chamber $\CW$
 of the subgroup \SU{N}. 
We obtain the solution to $\rho_i-\rho_{i+1}=+E$;
\begin{align}
\rho_i &=z_i= \frac{1}{2} (N+1-2i)E~,& z_i - z_j &= -(i-j)E~~,&z_i+z_j&= \{(N+1)-(i+j)\}E~~.
\end{align}
After we pick up residues at these points, the contribution $\CZ_{D_N}^{\CW}$ from the Weyl chamber $\CW$ to the partition function $\CZ_{D_N}$ is given by  an integral of $x$,
\begin{align}
\CZ_{D_N}^{\CW}
&=   \frac{1}{E^{N}} \frac{2}{N!2^{N-1}} (-1)^{N-1} N \left( \frac{E}{2}
 \right)^{N-1} \frac{1}{N} \left( \prod_{i<j,j-i\neq 1}
 \frac{(z_i-z_j)^2}{(z_i-z_j)^2-E^2}   \right)\cr
&\times \oint \frac{dx}{2\pi
 \sqrt{-1}} \prod_{i=1}^{N-1}\prod_{j=i+1}^{N}
 \frac{(z_i+z_j+2x)^2}{(z_i+z_j+2x)^2-E^2}\cr 
&=\frac{(-1)^{N-1} }{N!\cdot N\cdot   2^{N-1}} \oint \frac{d\nu}{2\pi \sqrt{-1}} 
 \frac{\nu - (N+1)}{\nu - 2N}\prod_{i=1}^{N-1} \frac{\nu - (2i+1)}{\nu -
 (2i)}
~~.
\label{eqn:intofD3} 
\end{align}
We have made a change of variable, 
$x \rightarrow \nu=2x/E+(N+1)$. 
Though $\CZ_{D_N}^{\CW}$ is not equal to the full $\CZ_{D_N}$, this calculation reveals the proper set of points which contribute to the partition function  $\CZ_{D_N}$.

In order to evaluate the full contribution, we must determine
 the multiplicities of contributions. 
Multiplicities originate from the transformation properties of the Weyl group. 
Permutations yield $N!$ terms and sign flip transformations bring the $2^m$ terms for some integer $m$. 
The determination of $m$ is the heart  of the problem of multiplicities and we will carry this out now. 
The integrand in (\ref{eqn:intofD3}) has an even number of poles.
A pole $x=w$ has a counterpart $-w$ in the $x$ plane.
The poles in $\nu$ plane in (\ref{eqn:intofD3}) are,
\begin{align}
N&=2p & \nu &=2,4,6,\cdots, 2p,2p+2,\cdots,4p,\cr
N&=2p+1 & \nu &= 2,4,6,\cdots, 2p,2p+4,\cdots,4p+2.
\label{eqn:polesd}
\end{align}
The poles $\nu =2,4,6,\cdots,2p$  
 in (\ref{eqn:polesd}) correspond to the minus poles in $x$ plane while the remainder to the plus. 
The contributions of plus poles and minus poles are equivalent
 if the orientation of the integration is considered. 
The integral (\ref{eqn:intofD3}) then consists of $p$ different terms.
This also shows that there are 
 diagrams which are not constructed from the fundamental root systems.
 
Now we return to the calculation of full contribution.
 We must take care of the orbits of these points under the Weyl transformation. 
The pole $\nu=2j$ represents a point,
\begin{align}
X_j^{(D)}&= (\phi_1,\cdots,\phi_N)=\left( (j-1)E,(j-2)E,\cdots,E,0,-E,\cdots,(j-N)E \right)~~.
\end{align}
This point $X_j^{(D)}$ is stable under the subgroup $H_j$ of the Weyl group whose order is $2^{j-1}$. 
The group $H_j$ is the isotropy group of the point $X_j^{(D)}$. 
There are $\# W/\# H_j=N! 2^{N-j}$ points which have the same residue as that at the point $X_j^{(D)}$.
 In order to compute the full contribution to $\CZ_{D_{N}}$,
 we must count the multiplicities of $p$ terms. 
Let us denote by $j$-type a point whose residue  is equal to that of $X_j^{(D)}$. 
The existence of the non-trivial isotropy group means that
 the point with $j \geq 2$  is on the boundary of the closure of a Weyl chamber. 
There are two $p$-type points in $\CW$. 
To evaluate the integral (\ref{eqn:intofD3}), we collect the residues
at $\nu = 2,4,6,\cdots, 2p$. 
If the rank $N$ is even $(N=2p)$, the factor $2^{N-j}$ runs from $2^{2p-1}$ to $2^{p}$.  
To calculate the minimal contribution,
 we find that the  multiplicity is equal to $2^{p-j}$ which is given by dividing $2^{2p-j}$ by $2^p$.  
If the rank $N$ is odd $(N=2p+1)$, the factor $2^{N-j}$ runs from $2^{2p}$ to $2^{p+1}$. 
In this case, the multiplicity can be read off as $2^{p-j}$  which is given by dividing $2^{2p-j+1}$ by $2^{p+1}$. Finally we must take the center $\BZ_2$ into account, which yields a factor $1/2$. 
We have carried out this evaluation for all $N$. 
The results are given as follows,
\begin{enumerate}
\item {\bf $N=2p$ case~~(\SO{4p})}\\
\begin{align}
\CZ_{D_{2p}}
&= \frac{ 1}{2^{2p}  (2p)} \sum_{j=1}^{p} 2^{2p-j}
\frac{2p-2j+1}{4p-2j} \frac{(2j-3)!!}{(2j-2)!!}\frac{(4p-2j-1)!!}{(4p-2j-2)!!}
~~,
\end{align}
\item {\bf $N=2p+1$ case~~(\SO{4p+2})}\\
\begin{align}
\CZ_{D_{2p+1}}
&= - \frac{1}{2^{2p+1} (2p+1)} \sum_{j=1}^{p} 2^{2p-j+1}
\frac{2p-2j+2}{4p-2j+2} \frac{(2j-3)!!}{(2j-2)!!}\frac{(4p-2j+1)!!}{(4p-2j)!!}
~~.
\end{align}
\end{enumerate}
Here $p\geq 1$. This results disagree with the previous works.
The validity of our results will be examined in Appendix A.

\section{B$_N$ and C$_N$}
\begin{figure}[h]
\centerline{\resizebox{15mm}{!}{\includegraphics{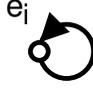}}}
\caption{(a) ; root $\xi_P \cdot e_i$,}
\label{fig:diag3}
\end{figure}

\hspace{12pt} Next we evaluate the partition functions for the root systems B$_{N}$
 and C$_N$. They are the root systems with respect to \SO{2N+1} and \USp{2N}. 
Their centers are $Z_{SO(2N+1)}=\{1\}$ and $Z_{USp(2N)}=\BZ_2$. 
The factors caused by these groups must be taken into account at the calculations of multiplicities.  
The partition functions for these root systems
 can be treated in parallel to the $D_N$ case in the last section. 
 The root system $B_N$ or $C_N$ has three types of roots;
 $\{\pm(e_i-e_j),i<j \}$, $\{\pm(e_i+e_j),i<j \}$ and 
 $\{\pm \xi_P \cdot  e_i \}$ ($P=B,C,~\xi_B=1, \xi_C=2$ ).
 The circles and the lines are the same as those for the D$_N$. 
Roots $\xi_P \cdot e_i$ are represented by cyclic lines introduced in Figure \ref{fig:diag3}.

We regard the cyclic line as a kind of loop subdiagram.
A connected diagram also has only one loop in these cases. 
So a connected diagram has a cyclic line or a loop $\divar{i}{j}{}$$\divbr{i}{j}{}$. 
The transformations on the diagrams can be defined and  
the reduction of the number of broken lines can be applied to these cases. 
The diagrams are transformed  into those with zero or one broken line. 
The partition function for the root system $P_N$ $(P=B,C)$ is,
\begin{align}
\CZ_{P_N} 
&= \frac{1}{E^{N}} \frac{\#Z_{P_N}}{N!2^{N}} \left(
 \prod_{\ell=1}^N  \oint \frac{d\phi_{\ell}}{2\pi
 \sqrt{-1}}  \right) \prod_{\alpha \in P_N} \frac{\langle \alpha, \phi \rangle}{\langle \alpha, \phi \rangle - E}~.
\label{eqn:intofB} 
\end{align}
where $\#Z_{B_N}=1$ and $\#Z_{C_N}=2$.
The variables, $x$ and $\rho_i$,  introduced in the case of $D_N$ are also valid,
\begin{align}
&x =\rho_0= \frac{1}{N}(\phi_1+\phi_2+\cdots \phi_N)~,&
\rho_i &= \phi_i - x,\cr
&\rho_1+\rho_2 + \cdots +\rho_N =0~, & J &=
 \frac{\partial(\rho_0,\rho_1, \cdots,\rho_{N-1})}{\partial(\phi_1, \phi_2 ,\cdots, \phi_N
 )} =\frac{(-1)^{N-1}}{N}~~,\\
\rho_i &=z_i= \frac{1}{2} (N+1-2i)E~,&&\cr
 z_i - z_j &= -(i-j)E~~,&
z_i+z_j&= \{(N+1)-(i+j)\}E~~.
\end{align}
We can perform the integrals of $\rho_i$ in the same manner.  
Then the contributions from  the Weyl chambers of the subgroups \SU{N} are,
\begin{align}
\CZ_{B_N}^{\CW}
&=\frac{(-1)^{N-1} }{N  2^{N+1}} \oint \frac{d\nu}{2\pi \sqrt{-1}} 
 \frac{(\nu-3)(\nu-(N+1))}{\nu(\nu-2(N+1))} 
\left( \prod_{i=2}^{N-1} \frac{\nu - (2i+1)}{\nu -2i} \right)
~~,
\label{eqn:intofB2} \\
\CZ_{C_N}^{\CW}
&=\frac{(-1)^{N-1}}{N  2^{N}} \oint \frac{d\nu}{2\pi \sqrt{-1}} 
 \frac{(\nu-N-1)}{(\nu-2N-1)} 
\prod_{i=1}^{N} \frac{(\nu-2i)}{(\nu-2i+1)}
~~,
\label{eqn:intofC2} 
\end{align}
where $\nu=2x/E+(N+1)$ again. 
For the root system B$_N$, the poles in the $\nu$-plane are
\begin{align}
N&=2p & \nu &=0,4,6,8,\cdots,2p,2p+2,\cdots,4p-2,4p+2,\cr
N&=2p+1 & \nu &= 0,4,6,8,\cdots,2p,2p+4\cdots,4p,4p+4.
\end{align}
For the root system C$_N$, the poles in the $\nu$-plane are
\begin{align}
N&=2p & \nu &=1,3,\cdots,2p-1, 2p+3,2p+5,\cdots, 4p-1,4p+1,\cr
N&=2p+1 & \nu &= 1,3,\cdots, 2p+1, 2p+3,\cdots, 4p+1,4p+3~.
\end{align}
These poles make pairs as the case of D$_N$.  
We must calculate the multiplicities to finish these calculations. 
For $B_{N}$ the pole  $\nu=2j$,  $(j=0,2,3,\cdots,p)$ represents a point,
\begin{align}
X_j^{(B)}&= (\phi_1,\cdots,\phi_N)=\left( (j-1)E,(j-2)E,\cdots,E,0,-E,\cdots,(j-N)E \right)~~.
\end{align}
For $C_{N}$ the pole  $\nu=2j-1$,  $(j=1,2,3,\cdots,[(N+1)/2] )$ represents a point,
\begin{align}
X_j^{(C)}&= (\phi_1,\cdots,\phi_N)=\left( (2j-3)E/2,(2j-5)E/2,\cdots,E/2,-E/2,\cdots,(2j-1-2N)E/2 \right)~~.
\end{align}
Orders of isotropy groups $H_j^{(B,C)}$ for $X_j^{(B,C)}$ ($1 \leq j$) are $2^{j-1}$
 and the order of $H_0^{(B)}$ for $X_0^{(B)}$ is $2^{N}$. 
These orders determine the multiplicities.
Carrying out the residue calculus, we obtain,
\begin{align}
\CZ_{B_{2p}}
&=\frac{1}{(2p)  2^{2p+1}}
\left[  2^{2p}
\frac{ (4p-1)!!}{(4p-2)!!} 
+  \sum_{j=2}^{p} 2^{2p-j+1} \frac{(2j-3)(2p-2j+1)}{2j(4p-2j+2)}
 \frac{(2j-5)!!}{(2j-4)!!} \frac{(4p-2j-1)!!}{(4p-2j-2)!!} 
  \right],\\
\CZ_{B_{2p+1}}
&=-\frac{ 1}{(2p+1)  2^{2p+2}}
\left[  2^{2p+1}
\frac{(4p+1)!!}{(4p)!!} 
+ \sum_{j=2}^{p}2^{2p-j+2}\frac{(2j-3)(2p-2j+2)}{2j(4p-2j+4)}
 \frac{(2j-5)!!}{(2j-4)!!} \frac{(4p-2j+1)!!}{(4p-2j)!!} 
  \right],\\
\CZ_{C_{2p}}
&=\frac{ 1}{(2p) 2^{2p+1}}
\sum_{j=1}^p2^{2p-j+1}\left[ 
\frac{(p-j+1)}{(2p-j+1)} \frac{(2j-3)!!}{(2j-2)!!}
 \frac{(4p-2j+1)!!}{(4p-2j)!!}  \right], \\
\CZ_{C_{2p+1}}
&= -\frac{ 1}{(2p+1) 2^{2p+2}}
\sum_{j=1}^{p+1}2^{2p-j+2}\left[ 
\frac{(2p-2j+3)}{(4p-2j+4)} \frac{(2j-3)!!}{(2j-2)!!}
 \frac{(4p-2j+3)!!}{(4p-2j+2)!!}  
  \right].
\end{align}
These expressions are valid for $p\geq 1$. 
We summarize the values of partition functions for B, C and D
 for small values of $p$ in Table \ref{tbl:results}.

\begin{table}\begin{center}
\begin{tabular}[h]{|l|ccccc|}
\hline
 $p$& 1&2&3&4&5\\
\hline
&&&&&\\
$D_{2p}$ &$\infrac{1}{8}$ & $\infrac{33}{256}$& $\infrac{125}{1024}$&
 $\infrac{29953}{262144}$& $\infrac{224577}{2097152}$  \\ 
&&&&&\\
$D_{2p+1}$& $\infrac{1}{8}$ & $\infrac{1}{8}$& $\infrac{483}{4096}$&
 $\infrac{3621}{32768}$& $\infrac{217705}{2097152}$  \\ 
&&&&&\\
$B_{2p}$ & $\infrac{3}{8}$ & $\infrac{71}{256}$& $\infrac{957}{4096}$&
 $\infrac{54195}{262144}$& $\infrac{196805}{1048576}$  \\ 
&&&&&\\
$B_{2p+1}$  & $\infrac{5}{16}$ & $\infrac{129}{512}$& $\infrac{7}{32}$&
 $\infrac{51501}{262144}$& $\infrac{754839}{4194304}$  \\ 
&&&&&\\
$C_{2p}$ & $\infrac{3}{16}$ & $\infrac{5}{32}$& $\infrac{1127}{8192}$&
 $\infrac{32589}{262144}$& $\infrac{478951}{4194304}$  \\ 
&&&&&\\
$C_{2p+1}$ & $\infrac{11}{64}$ & $\infrac{75}{512}$& $\infrac{4279}{32768}$&
 $\infrac{124765}{1048576}$& $\infrac{1844775}{16777216}$  \\ 
&&&&&\\
\hline
\end{tabular}
\caption{The absolute values of partition functions for root systems B,C and D.}
\label{tbl:results}
\end{center}\end{table}

\section{Accidental isomorphisms}
We have calculated the partition functions for all classical gauge groups. 
We must confirm our results. 
Let us check some of the well-known correspondence among lower dimensional Lie groups. 
Groups \SO{4}, \SO{5} and \SO{6}  are locally isomorphic to
 \SU{2}$\times$\SU{2}, \USp{4} and \SU{4}, respectively.  
One might think that the values of the partition functions in each pair should be equal.  
Let classical groups $G_1$ and $G_2$ be locally isomorphic to each other. 
The integration variables for $G_l$ are  $\phi_i^{(l)}$ $(l=1,2)$.  
The variables $\{ \phi_i^{(1)} \}$ do not coincide with $\{ \phi_i^{(2)} \}$ and we must consider the Jacobian.  
To clarify this argument we construct an explicit isomorphism between members of each pair. 
\subsection*{\SO{4} and \SU{2}$\times$\SU{2}}
The root system of \SO{4} is $D_2 $ and that of \SU{2}$\times$\SU{2} is $A_1 \oplus A_1$. 
Let us construct an  isomorphism which maps $\FD_2 =\{\pm(\phi_1-\phi_2), \pm(\phi_1+\phi_2)  \}$ onto 
$\FA_1\oplus \FA_1 =\{ \pm 2\varphi_1,\pm 2\varphi_2 \}$.

The isomorphism is given as,
\begin{align}
\begin{pmatrix}
\varphi_1\\ \varphi_2
\end{pmatrix} &= \frac{1}{2}
\begin{pmatrix}
1&-1\\
1&1
\end{pmatrix}
\begin{pmatrix}
\phi_1\\ \phi_2
\end{pmatrix}~~.
\end{align}  
Thus the Jacobian for the change of variables $(\phi_1,\phi_2) \rightarrow (\varphi_1,\varphi_2)$ is $2$. 
So the value of the integration for \SO{4} becomes $2(1/4 \times 1/4)$.  
Note that our prescription for the multiplicity has been defined
 such that the pre-factor $\# Z_{G} / \# W_{\Phi}$  is cancelled. 
The correspondence between \SO{4} and \SU{2}$\times$\SU{2} has been confirmed. 

\subsection*{\SO{5} and \USp{4}}
The root system of \SO{5} is B$_2$ and that of \USp{4} is C$_2$.
Let us construct an isomorphism which maps $\FB_2 =\{\pm(\phi_1-\phi_2), \pm(\phi_1+\phi_2),\pm \phi_1,\pm \phi_2 \}$ onto 
$\FC_2 =\{\pm(\varphi_1-\varphi_2), \pm(\varphi_1+\varphi_2),\pm 2\varphi_1,\pm 2\varphi_2 \}$.

 The isomorphism is given as,
\begin{align}
\begin{pmatrix}
\varphi_1\\ \varphi_2
\end{pmatrix} &= \frac{1}{2}
\begin{pmatrix}
1&1\\
-1&1
\end{pmatrix}
\begin{pmatrix}
\phi_1\\ \phi_2
\end{pmatrix}~~.
\end{align} 
Indeed under this change of variables, every element in $\FB_2$ maps to $\FC_2$. 
The Jacobian for the change of variables $(\phi_1,\phi_2) \rightarrow (\varphi_1,\varphi_2)$ is $2$. 
So the value of the integration for \SO{5} becomes $2(3/16)$.  
The correspondence between \SO{5} and \USp{4} has been confirmed. 

\subsection*{\SO{6} and \SU{4}}
The root system of \SO{6} is D$_3$ and that of \SU{4} is A$_3$.
\begin{align}
\FD_3 &= \{ \pm(  \phi_1-\phi_2),  \pm( \phi_1-\phi_3), \pm(  \phi_2-\phi_3),  \pm(  \phi_1+\phi_2),  \pm( \phi_1+\phi_3), \pm(  \phi_2+\phi_3 ) \},\\
\FA_3 &= \{ \pm(  \varphi_1-\varphi_2),  \pm( \varphi_1-\varphi_3), \pm(  \varphi_2-\varphi_3),\cr
&  \pm( 2 \varphi_1+\varphi_2+\varphi_3),  \pm( \varphi_1+2\varphi_2+\varphi_3), \pm( \varphi_1+ \varphi_2+2\varphi_3 ) \}.
\end{align}
 The isomorphism is given as,
\begin{align}
\begin{pmatrix}
\varphi_1\\ \varphi_2\\ \varphi_3
\end{pmatrix} &= \frac{1}{2}
\begin{pmatrix}
1&-1&1\\
1&1&-1\\
-1&1&1
\end{pmatrix}
\begin{pmatrix}
\phi_1\\ \phi_2 \\ \phi_3
\end{pmatrix}~~.
\end{align} 
Under this change of variables, every element of $\FD_3$ maps into that of $\FA_3$. 
The Jacobian for this change of variables 
$(\phi_1,\phi_2,\phi_3) \rightarrow (\varphi_1,\varphi_2,\varphi_3)$ is $2$. 
So the value of the integration for \SO{6} becomes $2(1/16)$.  
The correspondence between \SO{6} and \SU{4} has been confirmed. 

These demonstrations consolidate the correctness of our evaluations.

\section{Summary and discussions}

\hspace{12pt} We evaluated the partition functions for all classical gauge groups by using diagrammatic methods. 
The actions are given as  topologically twisted and 
 dimensionally reduced actions of  $d=4$ $\CN=1$ super Yang-Mills theories 
 with  classical (semi-)simple gauge groups. 
Our diagrammatic methods revealed multiplicities of poles which contribute equally. 
The multiplicities for points on the boundary of a Weyl chamber were given correctly. 
With similar manner, the partition functions for dimensionally reduced actions of 
 $d=6$ and $d=10$ $\CN=1$ super Yang-Mills theories might be evaluated. 

Our original motivation for carrying out matrix integrals for groups other than \SU{N}
 stems from an interesting class of USp and SO matrix models  
 \cite{Itoyama:1997gm,Ezawa:1998vm,Itoyama:1998uz,Itoyama:2005af}.
It may be that our evaluations are useful in investigating the dynamical generation of space-time, 
 which is so far examined in mean field approximations \cite{Nishimura:2001sx}.

{\bf \large Acknowledgement}
We are grateful to Yukinori Yasui and Takeshi Oota for useful comments. 
This work is supported by the 21 COE program 
 ``Constitution of wide-angle mathematical basis focused on knots" 
 and in part by the Grant-in Aid for scientific Research (No. 18540285) from Japan Ministry of Education.

\appendix

\section{Direct calculation for $D_2$, $B_2$ and $C_2$}
\hspace{12pt} We present the direct calculations for $D_2$, $B_2$ and $C_2$ in this appendix.

The root system $D_2$ is related to the Lie group \SO{4}.
The partition function is given by, 
\begin{align}
\CZ_{D_2} &= \frac{1}{E^2} \frac{2}{2! 2^2} \oint \frac{d\phi_1}{2\pi \sqrt{-1}}\oint \frac{d\phi_2}{2\pi \sqrt{-1}} \frac{(\phi_1-\phi_2)^2}{(\phi_1-\phi_2)^2-E^2} \frac{(\phi_1+\phi_2)^2}{(\phi_1+\phi_2)^2-E^2} ~~.
\end{align}
The  intersections of the lines $\phi_1\pm \phi_2=\pm E$ 
  contribute to the integral.

\begin{figure}[h]
\centerline{\resizebox{60mm}{!}{\includegraphics{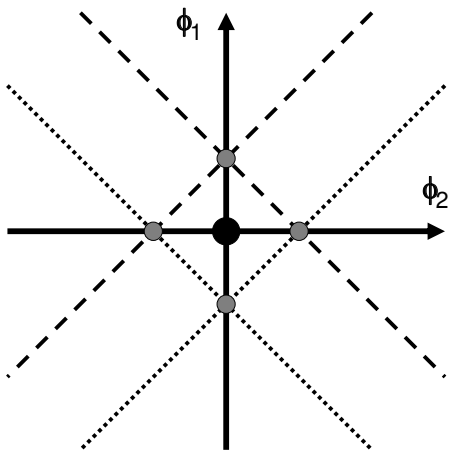}}}
\caption{The gray points in this figure contribute to the partition function of $D_2$. }
\label{fig:diagd2}
\end{figure}

The residues at all points take the same value. 
We calculate that at $(0,E)$ which is a intersection of $\phi_1-\phi_2=E$ and $\phi_1+\phi_2=E$. 
\begin{align}
&\res_{\phi_1-\phi_2=E} \left[ \frac{(\phi_1-\phi_2)^2}{(\phi_1-\phi_2)^2-E^2} \frac{(\phi_1+\phi_2)^2}{(\phi_1+\phi_2)^2-E^2}   \right]   =
\frac{E}{2} \frac{(2\phi_2 +E )^2}{4 \phi_2 (\phi_2 +E)} ~,\cr
&\res_{\phi_2=0}\frac{E}{2} \frac{(2\phi_2 +E )^2}{4 \phi_2 (\phi_2 +E)} =
\frac{E^2}{8}.
\end{align}

The root system $B_2$ is related to the Lie group \SO{5}.
The partition function is given by, 
\begin{align}
\CZ_{B_2} &= \frac{1}{E^2} \frac{2}{2! 2^2} \oint \frac{d\phi_1}{2\pi \sqrt{-1}}\oint \frac{d\phi_2}{2\pi \sqrt{-1}} \frac{(\phi_1-\phi_2)^2}{(\phi_1-\phi_2)^2-E^2} \frac{(\phi_1+\phi_2)^2}{(\phi_1+\phi_2)^2-E^2} \frac{\phi_1^2}{\phi_1^2-E^2}\frac{\phi_2^2}{\phi_2^2-E^2}  ~~.
\end{align}
The  intersections of the lines 
 $\phi_1\pm \phi_2=\pm E, \phi_i=\pm E$ contribute to the integral.

\begin{figure}[h]
\centerline{\resizebox{60mm}{!}{\includegraphics{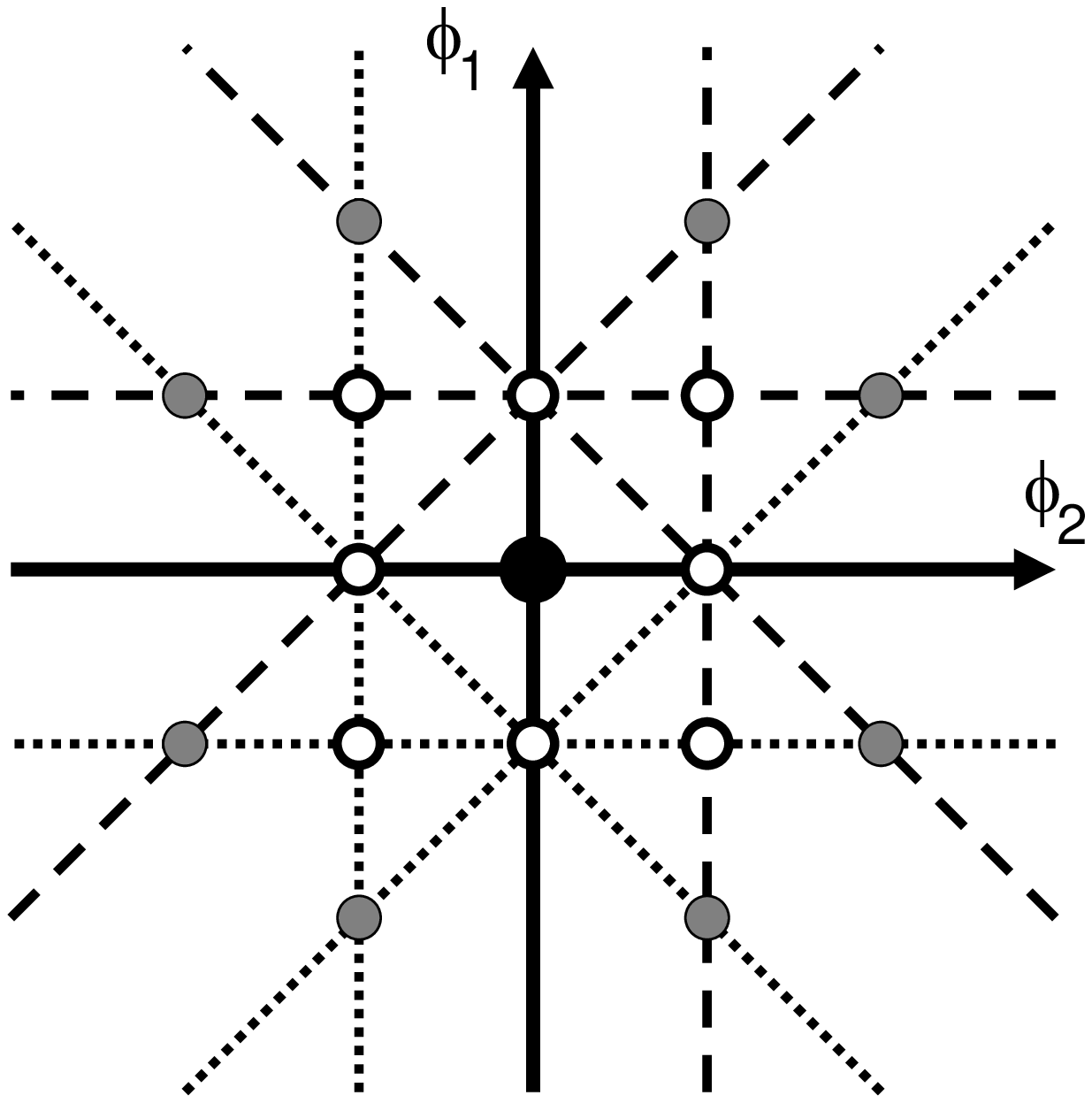}}}
\caption{The gray points in this figure contribute to the partition function of $B_2$ and the blank points do not.}
\label{fig:diagb2}
\end{figure}

The residues at all points also take the same value. 
We calculate that at $(2E,E)$ which is a intersection of $\phi_1-\phi_2=E$ and $\phi_2=E$. 
\begin{align}
&\res_{\phi_1-\phi_2=E} \left[ \frac{(\phi_1-\phi_2)^2}{(\phi_1-\phi_2)^2-E^2} \frac{(\phi_1+\phi_2)^2}{(\phi_1+\phi_2)^2-E^2} \frac{\phi_1^2}{\phi_1^2-E^2}\frac{\phi_2^2}{\phi_2^2-E^2}   \right]   \cr
&=\frac{E}{2} \frac{(2\phi_2 +E )^2}{4(\phi_2+ 2E)(\phi_2-E) } ~,\cr
&\res_{\phi_2=E} \frac{E}{2} \frac{(2\phi_2 +E )^2}{4(\phi_2+ 2E)(\phi_2-E) }=
\frac{3E^2}{8}.
\end{align}
One might pick up ``positive poles" upon $\phi_1$ integration,
 which are on the lines, $\phi_1\pm \phi_2=E$ and $\phi_1=E$.
We can find the solutions; $(2E,E),(-E,2E)$ and $(E,2E)$. Here we select the ``positives".
We  sum up the contributions from the three points.
Then the value of the partition function becomes $3\times2/2!2^2\times3/8=9/32$. 
If we divide the result $9/32$ by the order of the center $\BZ_2$,
 we obtain the result $9/64$ which is the same value in the previous works \cite{Krauth:1998xh,Kac:1999gw,Pestun:2002rr}.  
Our results in Table \ref{tbl:results} do not agree with this result. 

The root system $C_2$ is related to the Lie group \USp{4}.
The partition function is given by, 
\begin{align}
\CZ_{D_2} &= \frac{1}{E^2} \frac{2}{2! 2^2} \oint \frac{d\phi_1}{2\pi \sqrt{-1}}\oint \frac{d\phi_2}{2\pi \sqrt{-1}} \frac{(\phi_1-\phi_2)^2}{(\phi_1-\phi_2)^2-E^2} \frac{(\phi_1+\phi_2)^2}{(\phi_1+\phi_2)^2-E^2}\frac{4\phi_1^2}{4\phi_1^2-E^2}\frac{4\phi_2^2}{4\phi_2^2-E^2} ~~.
\end{align}
The  intersections of the lines $\phi_1\pm \phi_2=\pm E, 2 \phi_i=\pm E$ contribute to the integral.

\begin{figure}[h]
\centerline{\resizebox{60mm}{!}{\includegraphics{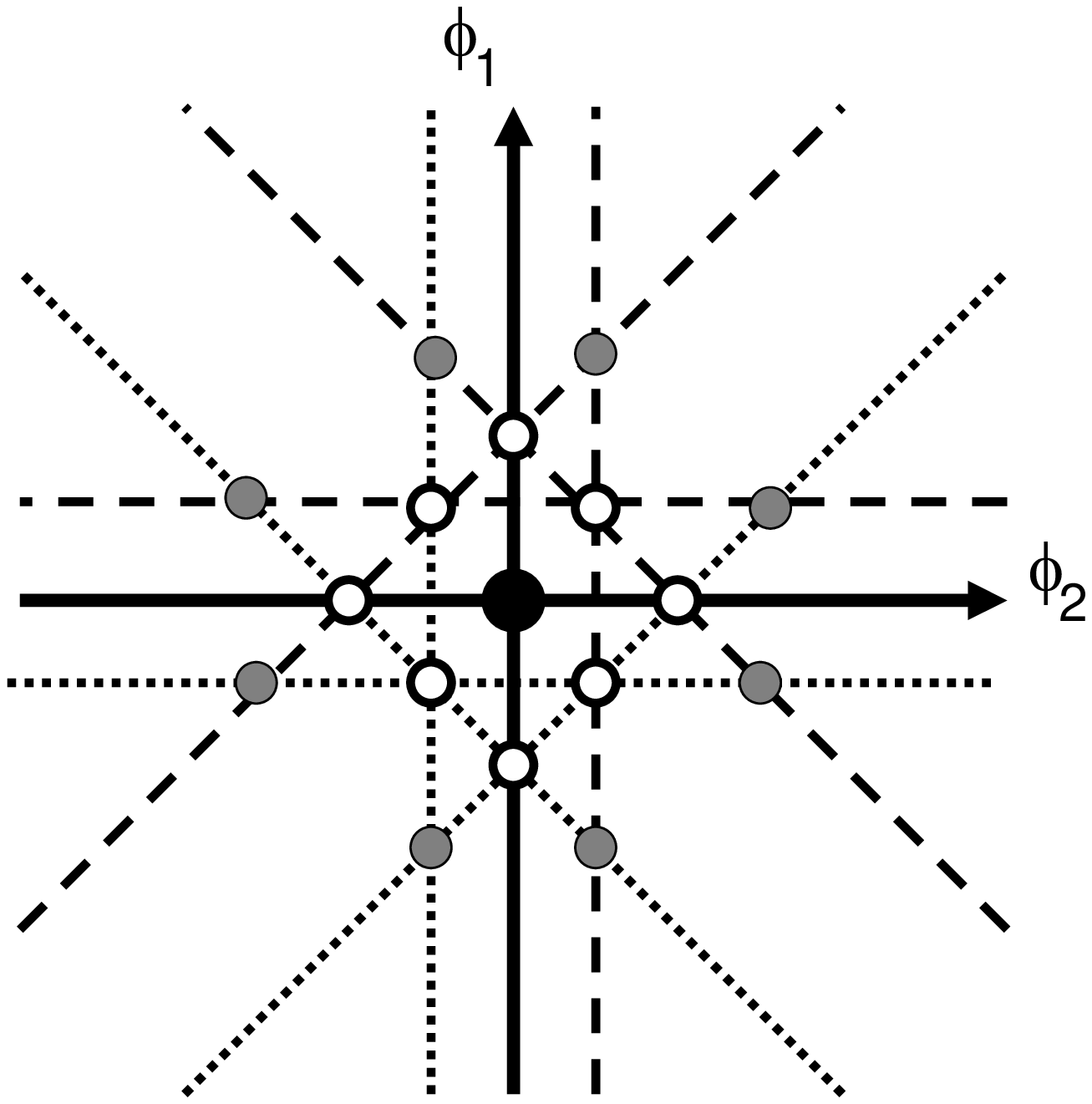}}}
\caption{The gray points in this figure contribute to the partition function of $C_2$ and the blank points do not.}
\label{fig:diagc2}
\end{figure}

The residues at all points take the same value. We calculate that at $(3E/2,E/2)$ which is a intersection of $\phi_1-\phi_2=E$ and $2\phi_2=E$. 
\begin{align}
&\res_{\phi_1-\phi_2=E} \left[ \frac{(\phi_1-\phi_2)^2}{(\phi_1-\phi_2)^2-E^2} \frac{(\phi_1+\phi_2)^2}{(\phi_1+\phi_2)^2-E^2} \frac{4\phi_1^2}{4\phi_1^2-E^2}\frac{4\phi_2^2}{4\phi_2^2-E^2}   \right]   \cr
&=\frac{E}{2} \frac{4\phi_2(\phi_2+E)}{(2\phi_2+3E)(2\phi_2-E)} ~,\cr
&\res_{\phi_2=E/2} \frac{E}{2} \frac{\phi_2(\phi_2+E)}{(\phi_2+3E/2)(\phi_2-E/2)}=
\frac{3E^2}{16}~.
\end{align}
These results support our calculation. 


\end{document}